# Temperature dependence of microwave and THz dielectric response in $Sr_{n+1}Ti_nO_{3n+1}$ (n=1-4)

D. Noujni[1], S. Kamba[1], A. Pashkin[1], V. Bovtun[1], J. Petzelt[1], A-K. Axelsson[2], N. McN Alford[3], P.L. Wise[3], I.M. Reaney[3]

[1]Institute of Physics, ASCR, Na Slovance 2, 18221 Prague 8, Czech Republic

[2]South Bank University, 103 Borough Road, London SE1 0AA, UK

[3]University of Sheffield, Department of Engineering Materials, Sheffield S1 3JD, UK

The microwave, near-millimetre and infrared (IR) dielectric response of $Sr_{n+1}Ti_nO_{3n+1}$ (n=1-4) Ruddlesden-Popper homologous series was studied in the temperature range 10 to 300 K. Remarkable softening of the polar optical mode was observed in $Sr_4Ti_3O_{10}$ and $Sr_5Ti_4O_{13}$ which explains the increase in microwave permittivity and dielectric loss upon cooling. However, both samples have a distinct content of $SrTiO_3$ dispersed between SrO layers. It is proposed therefore that the observed soft mode originates from the $SrTiO_3$ microscopic inclusions.

$Sr_{n+1}Ti_nO_{3n+1}$ belong to the class of Ruddlesden-Popper (RP) [1, 2] compounds, composed of *n* perovskite blocks of $SrTiO_3$ oriented along the [001] direction, separated and sheared by rock-salt type SrO layers. The number of perovskite layers controls the value of room temperature permittivity ε' of the system; $Sr_2TiO_4$ has ε'=37, $Sr_5Ti_4O_{13}$ shows ε'= 100 and the end member (n=∞) of the series, $SrTiO_3$, exhibits ε'= 290 [3,4]. Also microwave (MW) dielectric losses ε'' and temperature coefficient of resonance frequency TCF increases with n [3,4]. On the basis of the room temperature IR reflectivity and time-domain THz transmission (TDTS) spectroscopic studies [5], we explain this behaviour by the softening of the lowest frequency polar phonon mode. This paper combines the direct observation of the soft mode behaviour in $Sr_{n+1}Ti_nO_{3n+1}$ down to 10 K as well as MW dielectric properties down to 30 K.

IR reflectivity spectra were obtained using a FTIR spectrometer Bruker IFS 113v, TDTS measurements were performed using an amplified femtosecond laser system [5]. MW experiments were performed via a resonant cavity method, using the $TE_{01\delta}$ mode of dielectric resonator [6].

Fig. 1 shows temperature dependence of ε' and ε'' in $Sr_{n+1}Ti_nO_{3n+1}$ at frequencies near 3 GHz. Two characteristic features are seen: 1) the magnitude of ε' and its temperature dependence increases with n; the samples with n=3 and 4 exhibit temperature behaviour of ε' similar to incipient ferroelectric $SrTiO_3$. 2) ε'' shows non-monotonous behaviour which will be explained below.

FTIR reflectivity and TDTS spectra exhibit small temperature dependence for samples with n=1 and 2, corresponding to small changes of MW ε'(T). Samples with n=3 and 4 show remarkable changes predominantly in the frequency range below 100 $cm^{-1}$ (see Fig. 2a) due to mode softening (see Fig. 2b). FTIR reflectivity spectra were fitted together with TDTS spectra with a generalized four-parameter damped oscillator model [5]. Examples of the resulting ε'(ω) and ε''(ω) spectra of $Sr_5Ti_4O_{13}$, including MW and TDTS data are shown in Fig. 3. One can see that the MW ε'(T) is completely described by the polar phonon contribution and its increase upon cooling is the consequence of phonon softening. Temperature dependence of the soft mode (SM) frequency in $Sr_{n+1}Ti_nO_{3n+1}$(n=3,4,∞) is shown in Fig. 2b and one can see that it is very similar in all three samples. Only the dielectric strength of the SM increases with n and therefore the static $\varepsilon_0$ increases up to $\varepsilon_0 \approx$ $10^3$- $10^4$ in $SrTiO_3$ at low temperatures [7]. One can speculate that the increase of the number of perovskite layers in the samples leads to the rise of the strength of the SM. However, XRD studies of our samples revealed presence of $SrTiO_3$ (i.e. n=∞) inclusions in the samples with n=3 and 4 [3]. $Sr_4Ti_3O_{10}$ contains 10-20 % of n=∞ phase whereas $Sr_5Ti_4O_{13}$ has approximately twice this amount. Therefore, it is likely that the SM originates from microscopic inclusions of the n=∞ phase, implying that we are in fact measuring the effective dielectric response of a composite, including the polar phonon parameters [8]. Unfortunately, it has not proven possible to prepare single phase samples, even "epitaxial" thin films contain intergrowth defects whose concentration increases with n [9].

MW ε''(T) (Fig. 1b) needs special comment. Intrinsic losses in MW ceramics originate from multiphonon absorption and the dependence $\varepsilon''(T) \propto \omega T^m$ (m=1-2) should be valid [10]. Our results are completely different. The maximum in ε''(T) appears near 140 K in the n=1 sample due to weak extrinsic relaxation. We note that $\varepsilon''(T)_{max}$ even increased after annealing at 1000 °C. In the other three samples the low-temperature rise in ε''(T) has a different origin. It is probably connected directly with the SM softening (see Fig. 3b).

Finally we can conclude that the low-temperature dielectric behaviour in the MW range (as well as increasing TCF with n) can be explained with the polar SM which probably originates predominantly from the $SrTiO_3$ microscopic inclusions in our samples. In the present stage of our knowledge, we cannot distinguish the intrinsic phonon softening in the ideal $Sr_{n+1}Ti_nO_{3n+1}$ structure from that of $SrTiO_3$ inclusions.

The work was supported by the Czech grants (Projects No. A1010213, 202/01/0612, K1010104 and LN00A032).

Figure Captions:

Figure 1: Temperature dependent microwave dielectric permittivity (a) and loss (b) of $Sr_{n+1}Ti_nO_{3n+1}$ (n = 1-4) samples.

Figure 2: (a) Experimental FTIR reflectivity spectra of $Sr_4Ti_3O_{10}$ and $Sr_5Ti_4O_{13}$. (b) Temperature dependence of the lowest phonon mode frequencies of $Sr_{n+1}Ti_nO_{3n+1}$ (n = 3,4,∞). The data for $SrTiO_3$ are taken from [6].

Figure 3: Permittivity ε' (a) and dielectric loss ε'' (b) for $Sr_5Ti_4O_{13}$ calculated from the fits to the FTIR reflection and TDTS spectra. Experimental MW and TDTS data are marked with open and solid points, respectively.

Fig.1

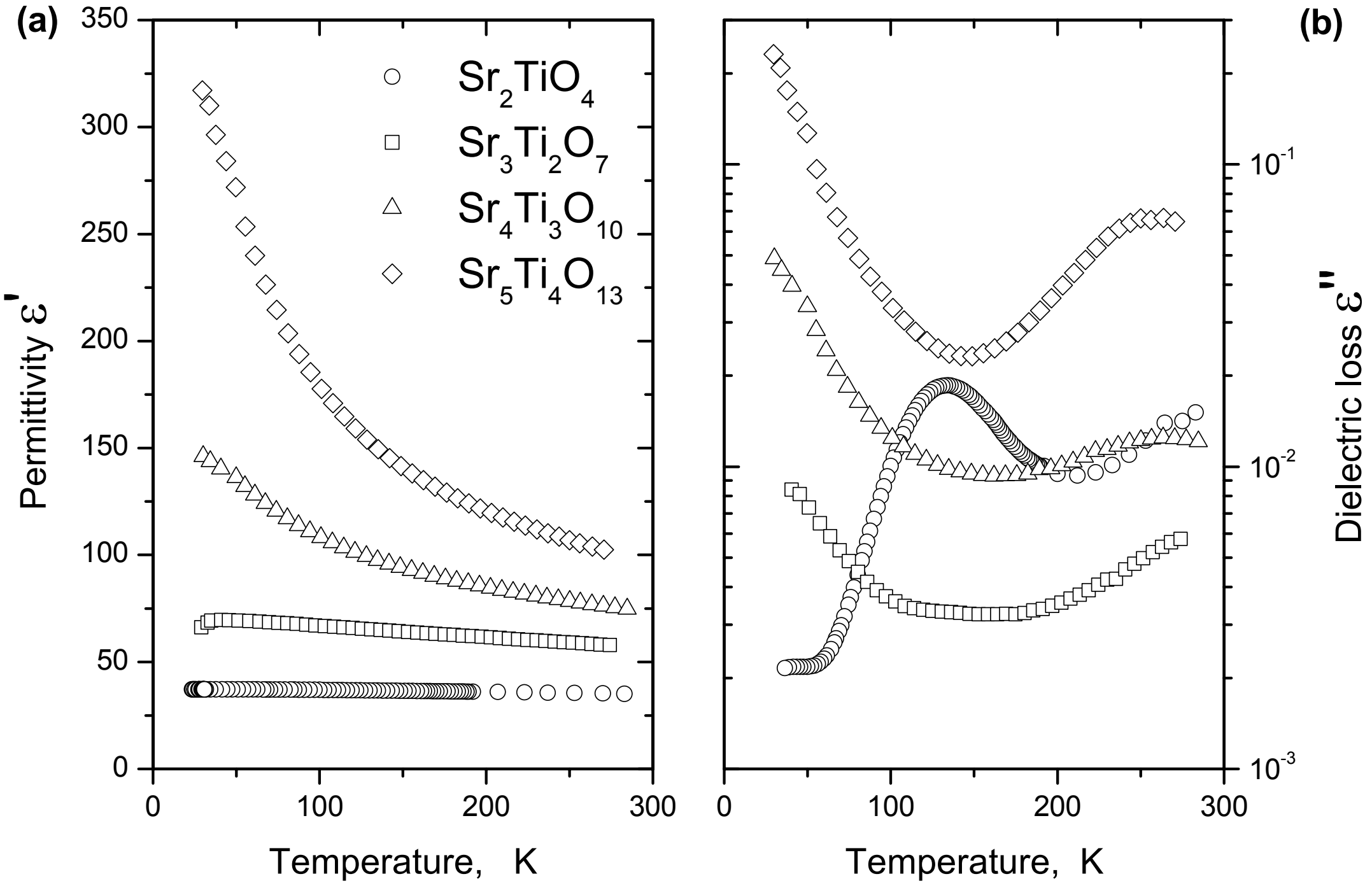

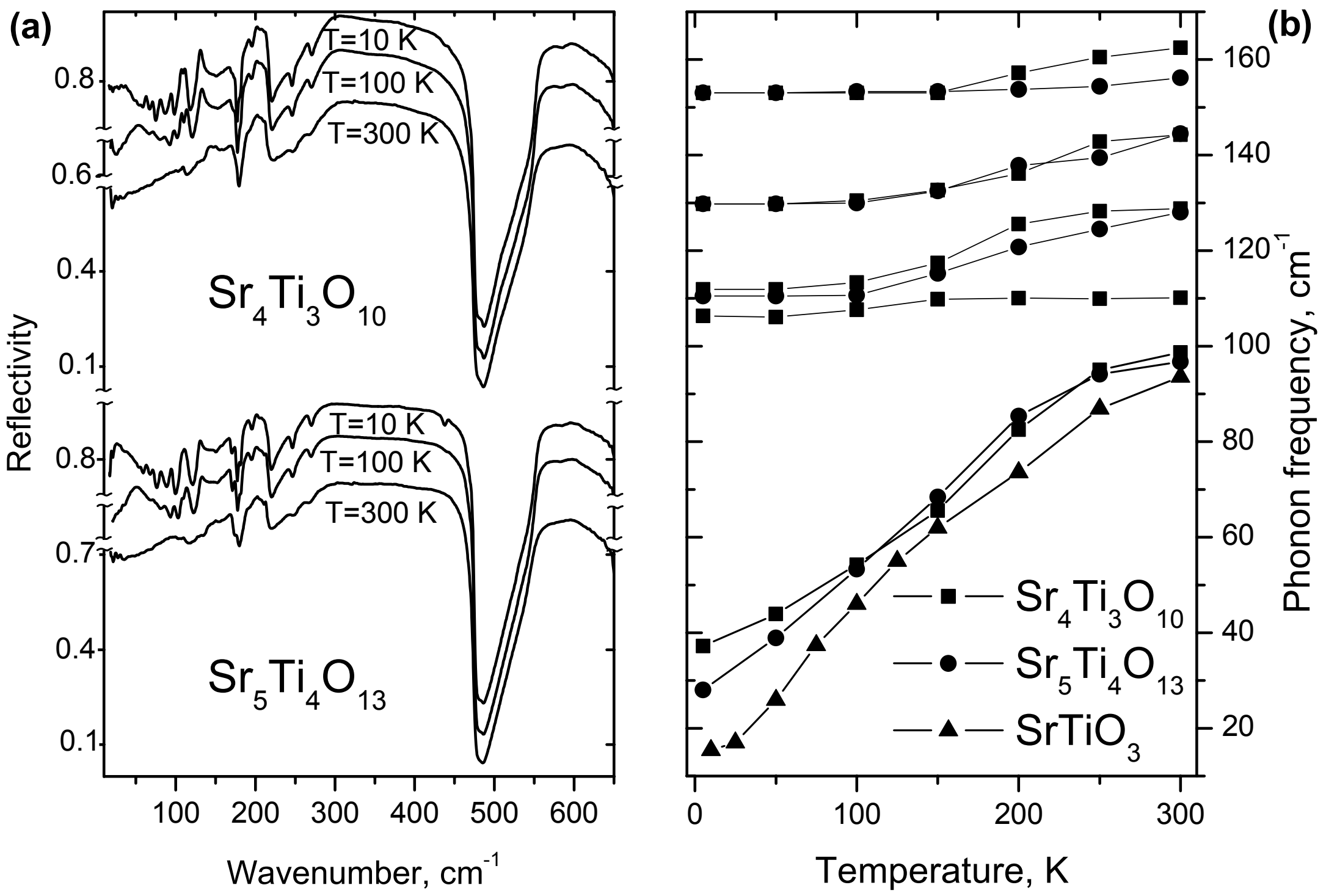

Fig.2

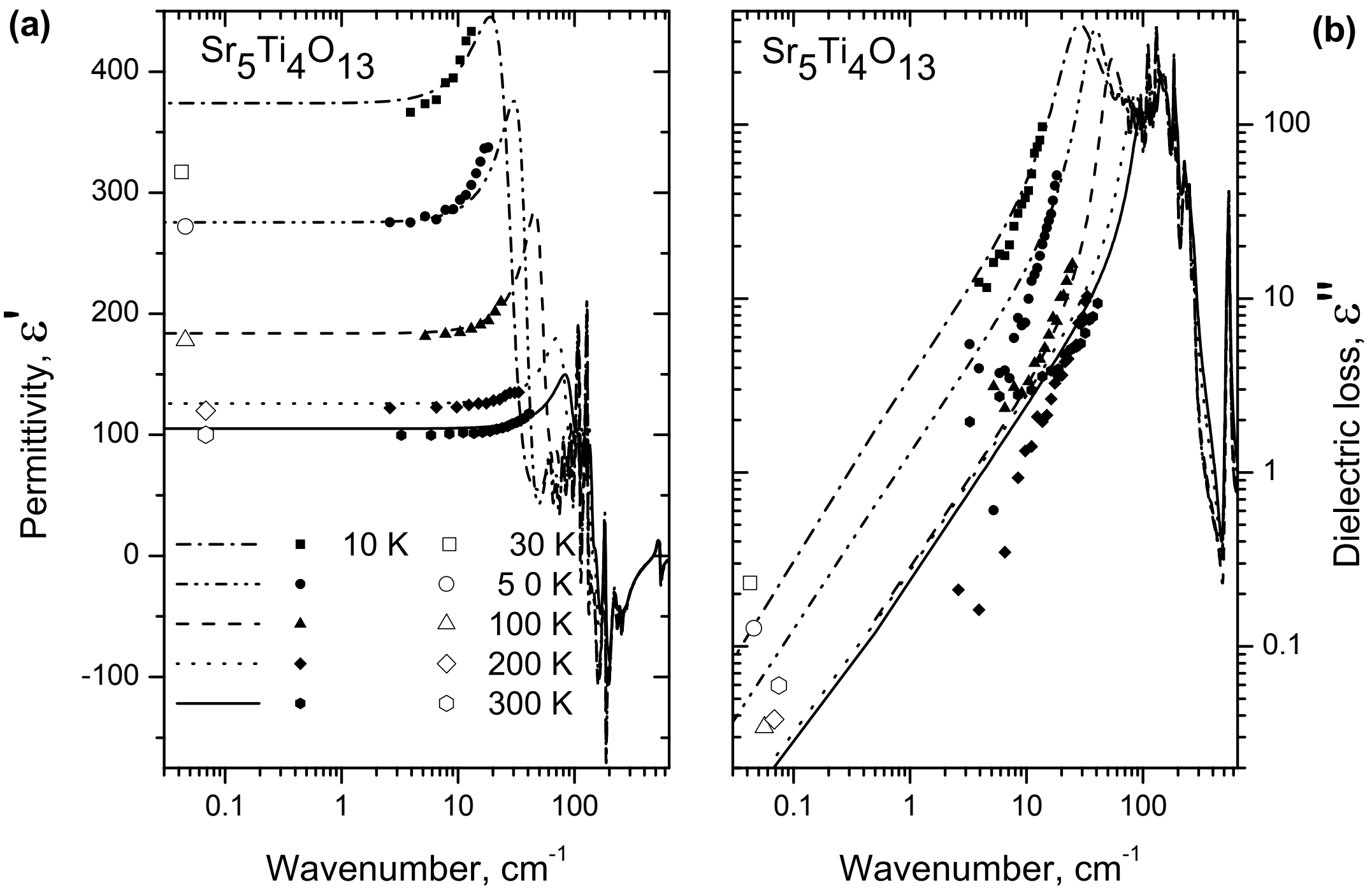

Fig.3